\documentclass[runningheads]{llncs}
\usepackage[T1]{fontenc}
\usepackage{graphicx}
\usepackage{booktabs}
\usepackage{amssymb}
\usepackage[normalem]{ulem} 
\usepackage{hyperref}
\usepackage{tikz}
\usetikzlibrary{arrows.meta, positioning}
\usepackage{float}
\begin{document}
\title{A Checklist to assess the energy and carbon impacts of ML/AI applications in Earth System Modeling}
\titlerunning{Checklist for Green AI in Earth System Modeling}
\author{Filippo Dainelli\inst{1}\orcidID{0000-0003-3278-7031} \and
Amirpasha Mozaffari\inst{1}\orcidID{0000-0001-6719-0425} \and
Marina Casta\~no\inst{1}\orcidID{0009-0007-6516-4320} \and
Aina Gaya i \`Avila\inst{1}\orcidID{0000-0001-6987-8901} \and
Llu\'is Palma Garcia\inst{1}\orcidID{0000-0002-3284-2152} \and
Alessio Melli\inst{1}\orcidID{0000-0002-8469-1624} \and
Oscar Dimdore-Miles\inst{1}\orcidID{0000-0002-3910-5606} \and
Amanda Duarte\inst{1}\orcidID{0000-0002-9340-958X}}
\authorrunning{Dainelli et al.}
\institute{Earth Department, Barcelona Supercomputing Center (BSC), Barcelona, Spain \\
\email{filippo.dainelli@bsc.es}
\thanks{Corresponding author.}
\thanks{F. Dainelli and A. Mozaffari contributed equally to this work.}}
\maketitle              % typeset the header of the contribution
\begin{abstract}
As machine learning and artificial intelligence find their way into nearly every aspect of climate, weather, and Earth system modeling, it is worth pausing to consider what our design decisions imply for the science and for the computational resources we consume. A growing body of literature addresses the ethical and sustainable development of ML/AI, yet translating these principles into day-to-day research practice remains a challenge as most of best practices are dispersed across multiple studies and commentaries. Here, we distill these discussions into a practical checklist that ML/AI and Earth system science practitioners can use to assess and reduce the environmental footprint of their own applications, organised around the successive stages of the model development pipeline. We complement the checklist with a selection of metrics drawn from the literature for estimating the energy consumption and carbon footprint of a project. For each question, we point to concrete examples and actionable suggestions from recent literature, aiming to bridge the gap between aspirational principles and the decisions researchers face at every stage of the development cycle.

\end{abstract}
%

%\begin{itemize}
%    \item Add concrete examples to support the claims and make the contribution more tangible
%    \item Merge duplicate or overlapping questions to reduce redundancy.
%    \item Prioritize the questions, giving the reader a clear sense of which ones are most pressing and should be addressed first.
%\end{itemize}

%\sout{- figure from npj}
%\sout{- examples FMs  foundation model utilization for fine tuning in ESM G2 (GraphCast examplese) -> PASHA}
%\sout{- most of the compute use is for R\&D G5 -> PASHA and  also comment that only reported final training is mentioned in ESM model (Pasha's paper)}
%\sout{- scalability law, Ferran thesis for G9 -> PASHA}
%\sout{- Aurora for G11 -> PASHA}
%\sout{- input variable selection and ablation studies for G3 -> FILIPPO}
%\sout{- some task comment about G4 -> FILIPPO}
%\sout{- show examples of emulators in AI systems G1 -> FILIPPO}
%\sout{- if able to find example for G8 or remove it -> FILIPPO}
%\sout{- add datasheet earth citations and comments in G10 -> FILIPPO
%    - https://journals.ametsoc.org/view/journals/bams/106/4/BAMS-D-24-0203.1.xml}
%
\section{Introduction}

Artificial Intelligence (AI) and Machine Learning (ML) models and algorithms have become increasingly skillful, efficient and powerful in performing different tasks across several sectors. However, much of this progress has been driven by increasingly large and computationally intensive models, resulting in more complicated architecture and more parameters to be trained \cite{boloncanedo2024greenai}. Schwartz et al.~\cite{schwartz2020green} define this trend as \textit{Red AI}, highlighting the focus on accuracy at the expense of efficiency in resource use, and instead advocate for \textit{Green AI}, which treats computational efficiency as a key evaluation criterion alongside accuracy in order to reduce AI's environmental footprint and lower the barriers to participation in the field.
This trend is also evident in Earth System Modeling (ESM); in recent years, the training and application of increasingly complex AI models has increased. This has been particularly evident for weather forecasting, where several ML-based models have been developed on different kinds of architectures \cite{pathak2022fourcastnet,bi2023panguweather,lam2023graphcast,li2023fuxiweather,kochkov2024neuralgcm,lang2024aifs,price2025gencast,lang2026aifscrps}. Rieutord et al.~\cite{rieutford2026carbonweather} provide a first quantitative estimation of the energy and carbon footprint of seven data-driven weather forecasting models, comparing their full life cycle (training plus inference) against the European Centre for Medium-Range Weather Forecasts (ECMWF) Integrated Forecast System (IFS), which is considered the most accurate General Circulation Model. Despite showing that in the long term the lower inference cost of the AI models compensates for their training cost, their work also highlights how the training phase still carries significant and highly variable costs that must be reported and, where possible, minimized. Recent works have demonstrated how simpler and lighter versions of AI models can achieve performance comparable to that of more complex versions \cite{valencia2025dataefficientensembleweatherforecasting,cachay2026ucastsurprisinglysimpleefficient}.
Moreover, ESM applications in AI have some specific amplifiers that directly affect their environmental impact. For instance, geographical extent, temporal coverage, spatial resolution, and variable choices directly translate into computational time and resources. As a result, the project's conceptual decisions already determine the environmental footprint from the outset. Additionally, in several applications benchmarking and validation are more extensive, as it is important to assess physical consistency in the outputs, avoiding reducing the analyses to performance metrics alone. These procedures increase the number of validation cycles. Another relevant factor is that traditionally in Earth System Science (ESS) AI models are trained from scratch rather than fine-tuning Earth System Foundation Models, which have also only emerged in recent years \cite{nguyen2023climax,schmude2024prithviwxc,bodnar2025aurora,watt-meyer2025ace2}.

The context highlights how the imperatives of \textit{Green AI} also apply to ESM. ML models are simultaneously tools for understanding the environment and consumers of the very resources that affect it. It is therefore fundamental to address these principles in the ESS community and provide practitioners with guidance that also considers specific aspects of the scientific domain, to help reduce the environmental impacts of AI/ML applications. Our proposal is a practical checklist structured around the different phases of an AI/ML project, so that relevant questions are raised at the most appropriate time, helping practitioners frame, develop and deploy their work with appropriate considerations on its environmental impact. The checklist is further supported by a selection of metrics extracted from the literature to help assess the energy consumption and carbon footprint of AI models, as well as references to open-source software that allow to keep track of these aspects throughout the development of their applications. Section~\ref{sec:checklist} presents the checklist questions divided by project phase. Section~\ref{sec:metrics} reports the collection of metrics that can be considered to assess the environmental impacts of an AI/ML project.

\section{Checklist} \label{sec:checklist}

\begin{figure}[t]
\centering
% ============================================================
% Navigation schema -- GREEN pillar only, adapted from the
% three-pillar flow chart of the full paper.
%
% Layout: vertical pipeline spine on the left (six stages),
% each linked to the checklist questions anchored to it.
%
% Requires in the main .tex preamble:
%   \usepackage{tikz}
%   \usetikzlibrary{arrows.meta, positioning}
%
% Numbering follows the checklist body of samplepaper-3:
%   Scoping G1-G2 | Data G3-G4 | Training G5-G8
%   Evaluation G9 | Reporting G10 | Deployment G11
% Stars follow the body: G3, G5, G8, G10.
% ============================================================
\begin{tikzpicture}[
  stage/.style={rectangle, rounded corners=2.5pt, draw=black!60, thick,
    fill=gray!12, minimum width=2.4cm, minimum height=0.8cm,
    align=center, font=\bfseries\small},
    gbox/.style={rectangle, rounded corners=2.5pt, draw=green!45!black,
    fill=green!8, minimum width=8.9cm, minimum height=1.05cm, align=left, font=\scriptsize,
    text width=8.5cm, inner sep=4pt},,
  arr/.style={-{Latex[length=2.3mm]}, thick, black!55},
  branch/.style={gray!45, thin}
]

\def\xS{0}        % stage column centre
\def\xG{6.05}     % question column centre
\def\vgap{4mm}  % vertical gap between question boxes

% ---------------- question boxes (chained top to bottom) ----------------
\node[gbox] (q1) at (\xG,0)
  {\textbf{G1} Justify net carbon cost vs.\ benefit\\
   \textbf{G2} Fine-tune rather than train from scratch};

\node[gbox] (q2) [below=\vgap of q1]
  {\textbf{G3}$\bigstar$ Investigate input features\\
   \textbf{G4} Match data granularity to task needs};

\node[gbox] (q3) [below=\vgap of q2]
  {\textbf{G5}$\bigstar$ Track all experiments\\
   \textbf{G6} Track carbon footprint with open libraries\\
   \textbf{G7} Use hyperparameter optimisation tools\\
   \textbf{G8}$\bigstar$ Profile scalability and hardware utilisation};

\node[gbox] (q4) [below=\vgap of q3]
  {\textbf{G9} Test simpler models and data subsets};

\node[gbox] (q5) [below=\vgap of q4]
  {\textbf{G10}$\bigstar$ Model/data card with full carbon footprint};

\node[gbox] (q6) [below=\vgap of q5]
  {\textbf{G11} Share weights, code and datasets};

% ---------------- stage boxes, aligned with their question row ----------
\node[stage] (s1) at (\xS,0 |- q1) {Scoping};
\node[stage] (s2) at (\xS,0 |- q2) {Data};
\node[stage] (s3) at (\xS,0 |- q3) {Training};
\node[stage] (s4) at (\xS,0 |- q4) {Evaluation};
\node[stage] (s5) at (\xS,0 |- q5) {Reporting};
\node[stage] (s6) at (\xS,0 |- q6) {Deployment};

% ---------------- connectors ----------
\foreach \s/\q in {s1/q1, s2/q2, s3/q3, s4/q4, s5/q5, s6/q6}{
  \draw[branch] (\s.east) -- (\q.west);
}

% ---------------- pipeline spine ----------
\foreach \a/\b in {s1/s2, s2/s3, s3/s4, s4/s5, s5/s6}{
  \draw[arr] (\a.south) -- (\b.north);
}

% ---------------- legend ----------
\node[font=\scriptsize, anchor=north west] at (s6.south west |- q6.south)
  [yshift=-3mm] {$\bigstar$ = core question, most pressing to address first};

\end{tikzpicture}
\caption{Navigation schema for the checklist. Each question is tagged to the
earliest stage at which it becomes actionable. $\bigstar$ marks core questions.}
\label{fig:schema}
\end{figure}
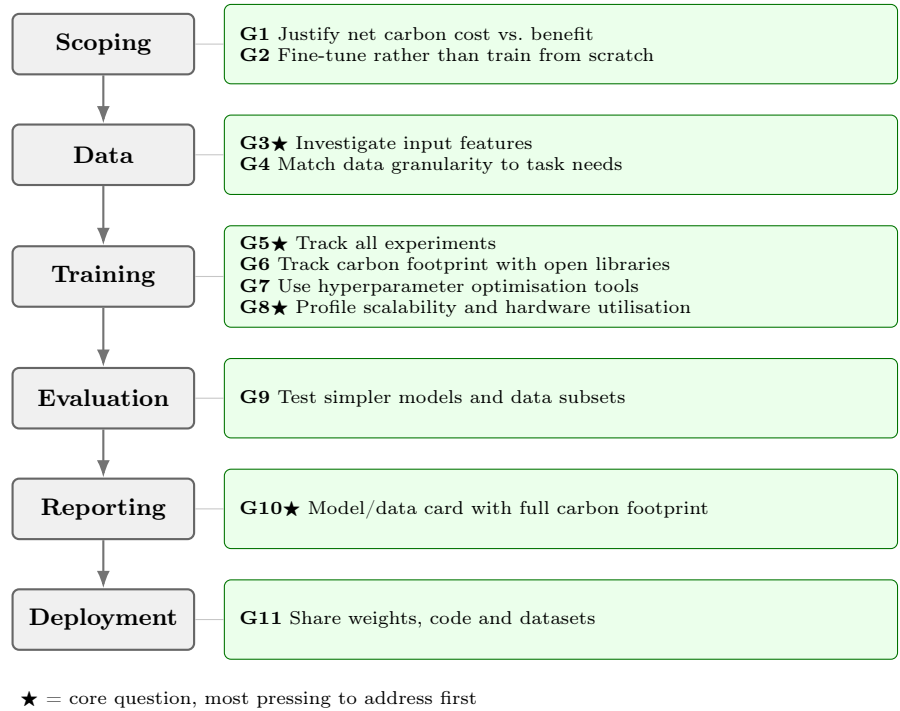

In building the checklist we aimed for a tool that is operational, so every question is assigned to a \textbf{Stage} which indicates the phase of the project at which it is most actionable. Several questions naturally span more than one stage; for those cases we tag the earliest stage at which the question should first be raised, since the cost of revisiting an issue grows the further downstream it is caught. The checklist is organised around six stages that follow the natural progression of an AI/ML project , summarised in Fig.~\ref{fig:schema}. \textit{Scoping} covers problem definition, model and architecture choice, and task framing; it is the stage at which the fundamental trade-offs between model complexity and resource consumption are first set, making it the most critical point for embedding Green AI principles. \textit{Data} encompasses collection, curation, splitting, and pre-processing; in ESM applications these steps often involve large-scale reanalysis or observational datasets, and decisions made here directly determine storage requirements and the computational cost of subsequent stages. \textit{Training} includes model fitting, experimentation, and optimisation, and typically represents the most energy demanding phase of the development cycle. \textit{Eval} covers validation, ablation studies, benchmarking, and interpretability checks; in ESM contexts this phase is more extensive than in traditional ML applications as physical consistency must be assessed alongside standard performance metrics. Moreover, in projects involving probabilistic forecasting, inference must be run several times to create a sufficient number of ensemble members. \textit{Reporting} addresses documentation, model and data cards, and the sharing of artefacts; transparent reporting of energy consumption and carbon footprint at this stage is essential for reproducibility. Finally, \textit{Deployment} considers release, accessibility, downstream use, and post-deployment monitoring; accounting for inference costs and the breadth of downstream use is necessary to obtain a complete picture of the environmental footprint of a model over its full operational lifetime.

The following list presents the checklist questions and actions organised by project stage, with most pressing questions denoted by $\bigstar$. To ease the self-assessment of the project, the questions and actions are reported in a self-assessment card, which can be found in Appendix~\ref{sec:A1}. \\

\textit{Scoping.}
\begin{itemize}
    \item[G1] If my goal is to substitute the numerical model with an ML/AI model (to improve inference speed, computation cost), have I estimated the net carbon impact of my model, and does the expected benefit justify the training and inference cost \cite{stern2025green,rieutford2026carbonweather}? Several AI emulators have been developed across ESM components. In weather forecasting, AIFS \cite{lang2024aifs} produces a 10-day global forecast in approximately 2 minutes 30 seconds on a single GPU, compared to the order of thousands of CPUs typically required by physics-based NWP models. In ocean modelling, Samudra \cite{dheeshjith2025samudra} emulates a state-of-the-art climate model ocean component roughly 150 times faster than the original numerical model. Similar speedups have been demonstrated also in seismic wave simulation, where Moseley et al.~\cite{moseley2020seismic} replace finite-difference simulation with an emulator running up to 549 times faster on GPU. These examples illustrate both the potential efficiency gains and the importance of weighing them against the upfront training cost.
    \item[G2] Do I need to train a model from scratch, or is there already a large, foundation model that I can use as the baseline and limit my training to adopting them, or fine-tune them \cite{zhu2026foundationsearthFM}? Aurora \cite{bodnar2025aurora} is pretrained once at a cost of 2.5 weeks on 32 A100 GPUs, then fine-tuned to air quality, ocean waves, cyclone tracks and 0.1° weather, outperforming the same architecture trained from scratch by 54\% on average across the CAMS targets. Reuse is cheaper still: Subich \cite{subich2025graphcastgdps} adapted GraphCast \cite{lam2023graphcast} to the Canadian operational analysis in 37 GPU-days, beating ECCC's operational forecast, and Lehmann et al. \cite{lehmann2025lightweightdecoders} extended Aurora to unseen hydrological variables with shallow decoders on frozen latents, at half the training time of full fine-tuning. The gain is conditional: accuracy tracks how closely the new variables correlate with those seen in pretraining. 
\end{itemize}

\textit{Data.}
\begin{itemize}
    \item[G3$\bigstar$] Do I need all the features I am training on, or only those that carry the strongest signal? Before or during \textit{Training}, feature selection algorithms can help reduce the dimensionality of the feature space, reducing computational time and resource usage while preserving performance \cite{guyon2003variableselection}. Dainelli et al.~\cite{dainelli20205xai-gpi} report how their feature selection schemes significantly reduce the input dimensionality without compromising the performance of their neural network for tropical cyclogenesis detection, while also enhancing model transparency and interpretability. Similarly, Fang et al.~\cite{fang2024variableselection} show that streamflow forecasting models trained on compact feature subsets matched or exceeded the skill of models using larger candidate pools. During \textit{Eval}, ablation studies and sensitivity analyses can confirm whether all features are truly necessary, and additionally help assess the robustness of the model by verifying that its performance is not dominated by one or two features \cite{verdecchia2022datacentricgreenai}. Grundner et al.~\cite{grundner2024ablation} ran a sequential feature selection ablation across 24 candidate inputs for a cloud cover parameterization and found that models using only 4-10 features matched the performance of the fill 24 feature network, showing most of the originally considered features were not significant. performed a sequential feature selection ablation study across 24 candidate inputs for a cloud cover parameterisation and found that models using only 4--10 features matched the performance of the full 24 features network, demonstrating that the majority of the originally considered features contributed no significant improvement to model performance.
    \item[G4] Have I carefully assessed the data granularity requirements of my task (spatial resolution, spatial domain, and temporal coverage) against what is strictly necessary? Once the project scope has been defined, working with unnecessarily detailed data adds computational cost without improving the target outcome. Several ESM applications do not require high data granularity to capture their target signal, as the following examples illustrate. ENSO-related sea surface temperature anomalies operate at basin scales that do not benefit from high spatial resolution data \cite{bellenger2014enso}. In terms of temporal resolution, MJO tracking is well captured at daily or 5-day resolution given its 30--60 day propagation timescale \cite{zhang2005mjo}. For seasonal to decadal variability applications, monthly means are sufficient to capture the relevant signal, avoiding the added cost of daily or sub-daily data \cite{doblas-reyes2013seasonal}. % Finally, regional climate studies can restrict the spatial domain to the area of interest, avoiding the computational cost of training on global data \cite{giorgi2015cordex}.
\end{itemize}

\textit{Training.}
\begin{itemize}
    \item[G5$\bigstar$] Do I keep track of all my experiments? Tracking is a key for traceability, experiments are linked to specific git commits preventing redundant runs caused by lost or untracked changes. Tools such as MLfow \cite{zaharia2018mlflow} and Weights \& Biases \cite{biewald2020wandb} support both experiment tracking and git integration. In addition, using experiment tracking tools helps with making informed decisions on training runs, avoids redoing them, and helps estimating the compute hours for the whole R\&D cycle rather than the final training run alone. The gap between the two is large: Denain and Wu \cite{epoch2026randdvstrainingcompute} estimate that final training runs accounted for only 9.6\% of OpenAI's 2024 R\&D compute spending, and 12.3\% and 22.6\% at two smaller developers, implying that a reported training cost understates the development footprint by roughly a factor of four to ten. ESM cannot yet be checked against this because the numbers are absent: of twelve frontier weather and climate models surveyed by Mozaffari et al. \cite{mozaffari2026riseofai}, three report no compute figure at all and the rest report only the final run. Tracking must therefore be enabled from the first experiment rather than reconstructed afterwards, and the cumulative project compute reported alongside the final run (G10).
    \item[G6] Am I tracking the carbon footprint of my training runs using existing open-source libraries, and accounting for the carbon intensity of the computing location? The same experiment can produce up to 30 times more CO$_2$ emissions depending on the energy grid used \cite{henderson2020systematic}; choosing the lower-carbon option requires no change to the model and has an immediate impact. Several free tools are available: CodeCarbon (\texttt{https://codecarbon.io}) \cite{lannelongue2021green} integrates directly into Python pipelines and estimates emissions based on hardware and location; CarbonTracker \cite{anthony2020carbontracker} similarly tracks and predicts the carbon footprint of deep learning training runs; the Green Algorithms calculator (\texttt{www.green-algorithms.org}) \cite{lannelongue2021green} supports a broader range of hardware and computing platforms beyond GPU/Python workflows, also offering location-aware estimates; GreenAlgorithms4HPC \cite{lannelongue2022hpc} is specifically designed for High Performance Computing (HPC) environments; and \cite{lannelongue2023primer} provides practical guidance on choosing the right estimation approach for a given setup. The Electricity Maps interactive tool (\texttt{app.electricitymaps.com/map}) provides a free real-time view of carbon intensity by country and region.
    %\item[G7] Have I tested whether my model really requires its full capacity, or whether training on a subset of my data is sufficient? If my model is going to be used for inference, would it benefit from pruning, quantizing, or distilling the model? \cite{li2025frequencyalignedknowledgedistillationlightweight,zhao2025moreunlockingspecializationtime,li2026distillingtimeseriesfoundation} For example, the original 3D Pangu weather paper \cite{bi2023panguweather} was later modified by another group to a 2D transformer yield to 20-30\% reduction in computation, positional bias was replaced by relative bias, reducing model size by 40\%, and a revised training procedure led to 30\% faster convergence without any further tuning \cite{to2024architectural}. Similarly, other studies show that training on a fraction of the available data not only yields similar performance but in some cases exceeds that of models trained on the full dataset \cite{valencia2025dataefficientensembleweatherforecasting}.
    \item[G7] Did I use a tool to optimize my hyperparameters instead of going for trial and error? Snoek et al.~\cite{snoek2012bayesianoptimization} describe hyperparameter tuning as "black art" that relies on expert experience and rules of thumb, adopting an automated optimization tools can substantially reduce the number of exploratory training runs need to reach a good configuration \cite{feurer2019automatedml}. Several optimization frameworks specifically designed for AI/ML tasks support automatic hyperparameter search, including Optuna \cite{akiba2019optuna} and KerasTuner \cite{omalley2019kerastuner}.
    \item[G8$\bigstar$]Have I profiled the scalability of my training and inference pipelines in terms of processor count, memory usage, and I/O performance? Understanding how my model scales with available hardware and inspecting the traces of its training and inference runs helps identify inefficiencies early, such as underutilised processors or memory bottlenecks or I/O stalls, that translate directly into wasted computational resources and unnecessary energy consumption. Wall-clock time alone will not reveal these inefficiencies; the informative quantity is sustained floating-point throughput as a fraction of the hardware's theoretical peak. Measured this way the spread is extreme: Yu et al. \cite{yu2026scalinglaws} report 37.2\% of H100 peak for Aurora \cite{bodnar2025aurora} against 1.03\% for GraphCast \cite{lam2023graphcast} and 0.022\% for SFNO \cite{bonev2023sfno}, meaning Aurora extracts approximately 36 times more arithmetic work per GPU-hour than GraphCast. Much of this underperformance is not architectural but rather an input-pipeline artefact: replacing the dataloader with synthetic data raises SFNO and Pangu \cite{bi2023panguweather} throughput by factors of 44 and 10 respectively. Mirabent Rubinat \cite{mirabent2026ai4land} report a case where a data layout requiring around ninety file opens per sample left the GPUs idle for most of each step; consolidating the data into fewer, larger files reduced it to nine, raised weak-scaling efficiency from 69\% to 86\% across eight ranks, and is projected to cut a full training run from 54 to 1.3 hours on bit-for-bit identical data.

\end{itemize}

\textit{Evaluation.}
\begin{itemize}
    \item[G9] Have I tested whether my model really requires its full capacity, or whether training on a subset of my data is sufficient? If my model is going to be used for inference, would it benefit from pruning, quantizing, or distilling the model? \cite{li2025frequencyalignedknowledgedistillationlightweight,zhao2025moreunlockingspecializationtime,li2026distillingtimeseriesfoundation} For example, the original 3D Pangu weather paper \cite{bi2023panguweather} was later modified by another group to a 2D transformer yield to 20-30\% reduction in computation, positional bias was replaced by relative bias, reducing model size by 40\%, and a revised training procedure led to 30\% faster convergence without any further tuning \cite{to2024architectural}. Similarly, other studies show that training on a fraction of the available data not only yields similar performance but in some cases exceeds that of models trained on the full dataset \cite{valencia2025dataefficientensembleweatherforecasting}.
\end{itemize}

\textit{Reporting.}
\begin{itemize}
    \item[G10$\bigstar$] Have I produced a model card or data card that includes energy consumption and carbon footprint information, and have I documented the full computational cost of the project, including all experimental runs and not just the final training run? Reporting only the final model's cost systematically underestimates the true environmental footprint of the R\&D cycle. Extending the model card framework proposed by Mitchell et al.~\cite{mitchell2019model} to include Green AI considerations supports transparency and enables the community to make informed decisions about model reuse and its associated environmental cost. For ESM practitioners, Connolly et al.~\cite{connolly2025datasheets} provide a datasheet format specifically adapted for Earth science datasets, which can serve as a practical starting point to which energy and carbon footprint information can be added.
   %\item[G11] Have I shared model weights, training code, and pre-processed datasets where possible, to avoid redundant retraining by other groups? Releasing artefacts converts a cost paid once into a resource the community reuses indefinitely. Aurora \cite{bodnar2025aurora} publishes code and weights openly as versioned releases under a persistent identifier, and Prithvi WxC \cite{schmude2024prithviwxc} was released on an open model hub together with a fine-tuned gravity-wave parameterisation, showing that downstream adaptations are worth sharing and not only the base model. The saving is concrete: the GraphCast \cite{lam2023graphcast} and Aurora adaptations discussed in G2 were possible only because the pretrained weights were public, and each would otherwise have required a full pretraining run to reproduce.
\end{itemize}

\textit{Deployment.}
\begin{itemize}
    \item[G11] Have I shared model weights, training code, and pre-processed datasets where possible, to avoid redundant retraining by other groups? Releasing artefacts converts a cost paid once into a resource the community reuses indefinitely. Aurora \cite{bodnar2025aurora} publishes code and weights openly as versioned releases under a persistent identifier, and Prithvi WxC \cite{schmude2024prithviwxc} was released on an open model hub together with a fine-tuned gravity-wave parameterisation, showing that downstream adaptations are worth sharing and not only the base model. The saving is concrete: the GraphCast \cite{lam2023graphcast} and Aurora adaptations discussed in G2 were possible only because the pretrained weights were public, and each would otherwise have required a full pretraining run to reproduce.
\end{itemize}

\section{Metrics} \label{sec:metrics}

To support researchers in assessing the checklist questions and actions outlined in Section~\ref{sec:checklist}, we report a non-exhaustive list of metrics that are helpful for evaluating the energy and carbon impact of an AI/ML project across the different project stages. The metrics selected are widely adopted in the Green AI literature and can be computed from information typically available to ESM researchers, without requiring a extensive background in computer science. We present them following the order in which they become relevant over the course of a project: first a metric that can be estimated at the design stage; then metrics measured during training, evaluation, and inference; after metrics that translate operational energy measurements into real world carbon estimates; and finally metrics that help assessing the broader context of the project. All metrics are summarised in Table~\ref{tab:metrics}. Since metrics are primarily used to document and communicate the environmental impact of a project, they are all associated with the reporting phase, in addition to any earlier stage at which they first become relevant.

A useful metric for the scoping and design phase of the project is the count of floating point operations (FLOPs), which helps comparing different candidate architectures before committing to an expensive training run. Being hardware agnostic, it cannot be directly translated into an energy or carbon estimate, but it helps identify which architectures are more computationally demanding.

Operational metrics, i.e. metrics computed during a run, are used to quantify the energy impact of a single experiment, during training, evaluation, and inference. Energy consumption ($kWh$) is the primary measurable environmental quantity at HPC scale, measuring the total electricity consumed by the CPU or GPU. Measuring it directly requires integrating a monitoring tool into the job, such as the NVIDIA Management Library (NVML) or CodeCarbon. CPU-hours or GPU-hours measure the wall-clock time of a single run multiplied by the number of processors used. They are a coarse native unit of measurement for HPC, directly available from job accounting. Researchers can obtain them without additional tooling, and they can be converted to $kWh$ by multiplying by the manufacturer's Thermal Design Power (TDP). TDP is the maximum amount of heat a chip is expected to generate under realistic sustained workloads, expressed in watts, and is typically used when direct power measurement is not available as it provides an easy and intuitive way to estimate energy consumption. However, it can overestimate it if GPU utilisation falls below 100\%, or underestimate it as TDP only accounts for the accelerator chip itself. GPU utilisation measures the fraction of the GPU actively used during a run, and helps refine the estimate of power consumption based on the actual utilisation of the unit, avoiding the overestimation that can arise when relying solely on TDP. On NVIDIA-based cluster nodes, GPU utilisation can be monitored with the \texttt{nvidia-smi} command, which reports both the GPU utilisation percentage and instantaneous power measurements, providing a more precise basis for energy consumption estimates than approaches based on TDP.

To convert energy consumption estimated from operational metrics into a full carbon estimate, specific metrics can be used. The standard unit for reporting carbon impact is $CO_2$eq, which is obtained by multiplying the energy consumption by the grid carbon intensity. The grid carbon intensity is a quantity derived from the local electricity grid that measures how much $CO_2$ equivalent is emitted per unit of electricity by a given power grid. It is typically expressed in $\frac{gCO_2eq}{kWh}$ and depends on the energy mix feeding that grid at a given time and location, meaning that the same energy consumption can result in vastly different carbon footprints depending on the location of the computing infrastructure. This is particularly significant for AI/ML applications in ESM, as research is often conducted within collaborative frameworks that provide access to multiple HPC server. Alongside this metric, the Power Usage Effectiveness (PUE) should also be used. PUE is the ratio of the total energy a data centre consumes to the energy actually delivered to the IT equipment; like the grid carbon intensity, it is an external factor dependent on the facility or location and not something measured during a run. The global average for PUE is $\sim$1.57 \cite{lannelongue2023greener}, but values specific to the facility can differ significantly and are not always publicly available. Grid carbon intensity and PUE should also be considered during the scoping and design phase, given their dependence on the location of the computing infrastructure. Additionally, to account for the full carbon impact of the project, embodied carbon should also be considered. Embodied carbon is the $CO_2$eq emitted to manufacture, transport, and eventually dispose of the hardware itself. Since this carbon is emitted at the moment the hardware is manufactured, it is typically amortised over the operational lifetime of the device. To compute it, one takes the total embodied footprint of a server or processor, divides it by its expected operational lifetime, and then allocates a share of that to the project in proportion to the time or fraction of the machine used. It is not something that can be measured directly from individual jobs, but rather relies on published life cycle assessments provided by manufacturers or reported in the literature.

Beyond operational measurements and their conversion to carbon estimates, other metrics can be used to account for impacts at the project level. The Pragmatic Scaling Factor (PSF) is a multiplier applied to a single final run to estimate the total footprint of the whole project, accounting for everything that happened before that final run, such as failed experiments, debugging, hyperparameter search, and ablation studies. The PSF is a deliberately rough multiplier assigned by the developers and is useful when the experiments performed prior to the production run were not systematically tracked or measured. Another metric is the energy-to-accuracy ratio, which is a normalized ratio that expresses the energy cost per unit of model performance gained. It combines a energy or carbon cost metric with a skill or accuracy metric into a single comparative figure. The energy-to-accuracy is most useful as a comparative measure when looking at to different model's configurations against each other, aiming at answering the question \textit{was the energy spent worth the performance gained?} and turning a cost metric into a cost-effectiveness metric. Another relevant metric is the data storage footprint, which measures the energy consumed by storage infrastructure to keep data on disk over time, including spinning disks, replication, and backup systems. Unlike metrics related to compute operations, this one is tied to data at rest and reflects the full data life cycle of the project. This metric is particularly relevant for ESM because of the large storage volumes and long retention periods typically associated with reanalysis and observational datasets in ESS. The data storage footprint is typically expressed as $\frac{kgCO_2eq}{TB \cdot yr}$ and has values on the order of 10 as reported by manufacturers \cite{lannelongue2023greener}. If the datasets are already available, the data storage footprint can already be estimated during the scoping phase.

\begin{table}[H]
  \caption{Green AI metrics relevant to ESM, with the project phases they are associated with, the checklist questions and actions they are connected to, and the references that introduce or apply them.\\ Stages: Sc = Scoping, Da = Data, Tr = Training, Ev = Evaluation, Re = Reporting, De = Deployment}
  \label{tab:metrics}
  \footnotesize
  \begin{tabular}{@{}l p{4cm} c c l@{}}
    \toprule
    \textbf{Metric} & \textbf{Description} & \textbf{Stages} & \textbf{Checklist} & \textbf{References} \\
    \midrule
    \multicolumn{5}{@{}l}{\textit{Core}} \\
    FLOPs & Hardware-agnostic count of floating-point operations per model output. & Sc·Re & G1, G9 & \cite{schwartz2020green,torned2023greenmlreview} \\
    Energy consumption (kWh) & Total electricity drawn by GPU/CPU/RAM during a compute job. & Tr·Ev·De·Re & G6, G10 & \cite{strubell2019energy,lannelongue2021green,henderson2020systematic} \\
    GPU-/CPU-hours & Processor count times wall-clock time. & Tr·Ev·De·Re & G5, G8, G10 & \cite{strubell2019energy,henderson2020systematic} \\
    TDP & Maximum sustained power draw specified by manufacturer. & Tr·Ev·De·Re & G6 & \cite{budenny2022eco2AI,lannelongue2023greener} \\
    GPU utilisation (\%) & Fraction of GPU capacity actively used; low utilisation (30--60\% typical) means TDP-based estimates overstate efficiency. & Tr·Ev·De·Re & G8 & \cite{henderson2020systematic,wu2022sustainableai} \\
    CO$_2$eq & Energy consumption multiplied by grid carbon intensity. & Tr·Ev·De·Re & G6, G10 & \cite{lacoste2019carbonemissions,schwartz2020green,lannelongue2021green} \\
    Grid carbon intensity & gCO$_2$eq per kWh from the local electricity grid. & Sc·Tr·Ev·De·Re & G6 & \cite{lacoste2019carbonemissions,henderson2020systematic,anthony2020carbontracker} \\
    PUE & Ratio of total energy to IT equipment energy (global avg.\ $\sim$1.57). & Sc·Tr·Ev·De·Re & G6 & \cite{strubell2019energy,lannelongue2021green,lannelongue2023greener} \\
    Embodied carbon & CO$_2$eq from hardware manufacturing amortised over server lifetime and proportioned to \% of utilization. & Tr·Ev·De·Re & G6, G10 & \cite{wu2022sustainableai,lannelongue2023greener} \\
    Pragmatic Scaling Factor (PSF) & User-estimated multiplier applied to single-run emissions to account for debugging and hyperparameter sweeps across a project. & Re & G5, G10 & \cite{lannelongue2021green,bannour2021carbonnlp} \\
    Energy-to-accuracy ratio & kWh consumed per unit of accuracy/skill improvement. & Ev·De·Re & G9 & \cite{peykani2026greenaireview,schwartz2020green} \\
    Data storage footprint & Lifecycle CO$_2$eq of storing datasets ($\sim$10 kgCO$_2$e/TB/yr). & Sc·Da·Re & G4 & \cite{lannelongue2023greener} \\
    \bottomrule
  \end{tabular}
\end{table}

\section{Conclusion and Outlook}

As ML/AI methods become standard across climate, weather, and ESM, the resources spent developing and running them can no longer be treated as an afterthought. In this work we distilled the growing Green AI literature into a practical, stage-based checklist tailored to ESS, complemented by a curated set of metrics for estimating a project's energy consumption and carbon footprint. Our aim is not to propose new principles but to make existing ones actionable: by attaching each question to the stage at which it is most consequential, we help practitioners raise the right considerations early, when acting on them is cheapest. This matters especially in ESS, where choices of resolution, domain, temporal coverage, and ensemble size act as amplifiers that translate directly into compute and carbon.
The checklist is deliberately non-exhaustive and meant as a living tool that groups can adapt and the community can extend as practices evolve. The present version centres on environmental sustainability; future iterations could broaden toward the related dimensions of equity, fairness, and trustworthiness. We also see value in shared reporting norms so that a model's environmental cost becomes as routine to report as its skill. ML models in ESS are both instruments for understanding the environment and consumers of the resources that shape it; keeping both in view is, we hope, a step toward research practice that is effective and responsible at once.

\section*{Declaration on Generative AI}

Generative AI was used solely for polishing the language of this manuscript; it was not used for generating content or text nor for performing any analysis. All AI-assisted text was reviewed and verified by the authors, who take full responsibility for the accuracy and integrity of the manuscript.

\section*{Acknowledgements}

\begin{sloppypar}
AM acknowledges the Grant JDC2023-051208-I, funded by MICIU/AEI/10.13039/501100011033 and, as appropriate, by ``ERDF A way of making Europe'', by ``ERDF/EU'', by the ``European Union'', or by the ``European Union NextGenerationEU through PRTR''. AD acknowledges her AI4S fellowship within the ``Generación D'' initiative by Red.es, Ministerio para la Transformación Digital y de la Función Pública, for talent attraction (C005/24-ED CV1), funded by NextGenerationEU through PRTR. This work received funding from the European Union's Horizon Europe Framework Programme through the project CONCERTO (Grant Agreement 101185000). This work, as part of TerraDT -- Digital Twin of Earth System for Cryosphere, Land Surface and related interactions, has received funding from the European Union's Horizon Europe Framework Programme (HORIZON) under Grant Agreement no.\ 101187992. This work was supported by the European Union's Horizon Europe research and innovation programme under grant agreement No 101214398 (ELLIOT).
\end{sloppypar}

\begin{credits}
\subsubsection{\discintname}
The authors have no competing interests to declare that are relevant to the content of this article.
\end{credits}

%
% ---- Bibliography ----
%
\bibliographystyle{splncs04}
\bibliography{sample-ceur}

\appendix

\section{Self-Assessment Card} \label{sec:A1}

The following card is designed as a practical companion to the checklist in Section~\ref{sec:checklist}. For each question, mark one of four responses: \textit{Done}, \textit{Aware} (recognised but not yet addressed), \textit{No} (not considered), or \textit{N/A}. A standalone fillable version in Markdown format is available as supplementary material.

\begin{table}[H]
  \caption{Self-Assessment Card. $\bigstar$ marks core questions; tick one of \textit{Done}, \textit{Aware}, \textit{No}, \textit{N/A} per row.}
  \label{tab:green-selfassessment}
  \footnotesize
  \begin{tabular}{@{}c p{6cm} c c c c c c@{}}
    \toprule
    \textbf{ID} & \textbf{Question} & \textbf{Stage} & \textbf{Core} & \textbf{Done} & \textbf{Aware} & \textbf{No} & \textbf{N/A} \\
    \midrule
    G1  & Justify net carbon cost vs.\ benefit          & Scoping    &            & $\square$ & $\square$ & $\square$ & $\square$ \\
    G2  & Fine-tune instead of training from scratch    & Scoping    &            & $\square$ & $\square$ & $\square$ & $\square$ \\
    G3  & Investigate input features                    & Data       & $\bigstar$ & $\square$ & $\square$ & $\square$ & $\square$ \\
    G4  & Match data granularity to task needs          & Data       &            & $\square$ & $\square$ & $\square$ & $\square$ \\
    G5  & Track all experiments                         & Training   & $\bigstar$ & $\square$ & $\square$ & $\square$ & $\square$ \\
    G6  & Track carbon footprint with open libraries    & Training   &            & $\square$ & $\square$ & $\square$ & $\square$ \\
    G7  & Use hyperparameter optimization tools         & Training   &            & $\square$ & $\square$ & $\square$ & $\square$ \\
    G8  & Scalability and utilisation analysis          & Training   & $\bigstar$ & $\square$ & $\square$ & $\square$ & $\square$ \\
    G9  & Test simpler model and data subsets           & Evaluation &            & $\square$ & $\square$ & $\square$ & $\square$ \\
    G10 & Model/Data card with full carbon footprint    & Reporting  & $\bigstar$ & $\square$ & $\square$ & $\square$ & $\square$ \\
    G11 & Share weights, code, and datasets             & Deployment &            & $\square$ & $\square$ & $\square$ & $\square$ \\
    \bottomrule
  \end{tabular}
\end{table}

\end{document}